\documentclass[fleqn,11pt]{wlscirep}
\usepackage[utf8]{inputenc}
\usepackage[T1]{fontenc}
\usepackage{siunitx}  
\usepackage[version=4]{mhchem}

\usepackage{soul}

\usepackage{lineno}

\DeclareSIUnit\electron{\mathrm{e^-}}
\DeclareSIUnit\angstrom{\text{Å}}
 
\title{Deep generative priors for robust and efficient electron ptychography}

\author[1,*]{Arthur R. C. McCray}
\author[2]{Stephanie M. Ribet}
\author[3]{Georgios Varnavides}
\author[1,$\dagger$]{Colin Ophus}
\affil[1]{Department of Materials Science and Engineering, Stanford University, Stanford, CA 94305, USA}
\affil[2]{National Center for Electron Microscopy, Molecular Foundry, Lawrence Berkeley National Laboratory, Berkeley, CA 94720, USA}
\affil[3]{Department of Imaging Physics, Delft University of Technology, Lorentzweg 1, 2628 CJ Delft, the Netherlands}

\affil[*]{amccray@stanford.edu}
\affil[$\dagger$]{cophus@stanford.edu}

\begin{abstract} 
Electron ptychography enables dose-efficient atomic-resolution imaging, but conventional reconstruction algorithms suffer from noise sensitivity, slow convergence, and extensive manual hyperparameter tuning for regularization, especially in three-dimensional multislice reconstructions. 
We introduce a deep generative prior (DGP) framework for electron ptychography that uses the implicit regularization of convolutional neural networks to address these challenges. 
Two DGPs parameterize the complex-valued sample and probe within an automatic-differentiation mixed-state multislice forward model. 
Compared to pixel-based reconstructions, DGPs offer four key advantages: (i) greater noise robustness and improved information limits at low dose; (ii) markedly faster convergence, especially at low spatial frequencies; (iii) improved depth regularization; and (iv) minimal user-specified regularization. 
The DGP framework promotes spatial coherence and suppresses high-frequency noise without extensive tuning, and a pre-training strategy stabilizes reconstructions. 
Our results establish DGP-enabled ptychography as a robust approach that reduces expertise barriers and computational cost, delivering robust, high-resolution imaging across diverse materials and biological systems.

\end{abstract}
\begin{document}

\flushbottom
\maketitle

Electron ptychography has emerged as a powerful technique for high-resolution and dose-efficient imaging across diverse sample types. 
In materials science, ptychography routinely achieves sub-angstrom resolution, while for beam-sensitive and biological samples it enables imaging at or near the theoretical resolution limits~\cite{rodenburg2008,pennycook2015, huang2023a, nellist1998, chen2021}. 
Ptychography combines information from overlapping diffraction disks and leverages redundancy between scan positions to recover the complex-valued electron exit wave, overcoming key limitations of conventional scanning transmission electron microscopy (STEM)~\cite{maiden2013}. 
Recent advances in detector technology, computational algorithms, and data acquisition strategies have enabled ptychographic reconstructions of increasingly diverse systems, including two-dimensional (2D) heterostructures, crystalline defects, and frozen-hydrated proteins~\cite{jiang2018, zhu2025, kucukoglu2024}. 
Nonetheless, ptychographic reconstructions remain computationally demanding and require considerable expertise to obtain high-quality results. 

Multiple algorithms exist to reconstruct electron ptychography datasets~\cite{maiden2009, rodenburg1992, varnavides2023, clark2025, sanchez-santolino2025}. 
Direct methods are popular for their computational efficiency. Recent machine learning-based direct methods offer improvements via physics-informed neural networks and end-to-end learning frameworks~\cite{chang2023, babu2023, seifert2021, friedrich2023, cherukara2020}, though these often require large training datasets or lack interpretability. 
Iterative methods are popular because they simultaneously reconstruct both the sample object and electron probe, allowing ptychography to overcome aberrations in the probe-forming optics~\cite{maiden2009}. 
While effective, pixel-based iterative methods -- which reconstruct the object and probe on equispaced grids -- can suffer from slow convergence, sensitivity to noise, and susceptibility to artifacts.
This is particularly limiting when imaging samples with weak scattering contrast, sparse features such as nanoparticles, or when operating in dose-limited regimes~\cite{zhou2020b, chen2020b, pelz2017}. 
Automatic differentiation (AD) increases flexibility and enables joint optimization of auxiliary parameters~\cite{kandel2019, lee2025, gilgenbach2025}, but AD does not resolve the underlying challenges inherent to pixelated gradients.

Inverse multislice ptychography, which accounts for multiple scattering by decomposing the sample into thin slices along the beam direction~\cite{maiden2012}, adds degrees of freedom that exacerbate these issues and can yield nonphysical reconstructions without either extensive regularization or combining multiple projection directions~\cite{ribet2024, pelz2023}.
The choice of regularization is critical but nontrivial.
Common approaches include total variation (TV) penalties, sparsity constraints in transformed domains, and Fourier-space depth regularization~\cite{zhang2021b, moshtaghpour2025, wakonig2020}. 
However, these methods introduce additional hyperparameters that must be carefully tuned for each dataset, and aggressive regularization can suppress genuine structural features.

Deep generative priors (DGPs), also known as deep image priors~\cite{ulyanov2018}, offer an alternative paradigm that leverages the implicit regularization of convolutional neural network (CNN) architectures without pre-training on external datasets with known ground-truths. 
Originally developed for image restoration, DGPs have demonstrated that the structure of a CNN can serve as a powerful prior that favors natural, structured outputs over noise. 
This results from the inductive biases inherent to convolutional architectures, such as translation invariance and spatial hierarchies, which tend to generate outputs with spatial coherence and characteristic low-frequency structure. While initially applied to 2D images, the DGP framework naturally extends to higher-dimensional data and has been successfully applied to various inverse problems in computational imaging~\cite{gandelsman2019, zhou2020, liu2019, mccray2024}.
These methods have seen limited use in coherent diffractive imaging, and applying them to ptychography introduces distinct challenges and opportunities.

In this work, we develop a DGP-enabled AD framework for electron ptychography that addresses key limitations of conventional reconstruction algorithms. 
Two separate DGPs generate the complex-valued sample object and probe, embedding them within a differentiable mixed-state multislice forward model. 
This approach provides four primary advantages: (i) enhanced noise robustness, enabling lower-dose reconstructions with improved information limits; (ii) accelerated convergence for challenging samples, particularly those with significant low spatial frequency components; (iii) physically plausible 3D multislice reconstructions of thin specimens; and (iv) the avoidance of user-specified regularization terms. 
We introduce a pre-training strategy that initializes the DGPs from conventional reconstructions, ensuring stable optimization despite the high dimensionality of the joint problem. 
We apply our method to multiple experimental datasets spanning diverse materials and biological systems and imaging conditions, showing consistent improvements over conventional AD-based pixelated reconstructions.
We have integrated these methods into the open-source Python package quantEM to make them broadly accessible. 
Together, these results establish DGP-enabled ptychography as a robust, efficient, and versatile approach to ptychographic imaging across a broad range of experimental conditions and sample types.

\section*{Results}
\subsection*{Pre-training for stable reconstructions}
\begin{figure}[htb]
\centering
\includegraphics[width=1.0 \linewidth]{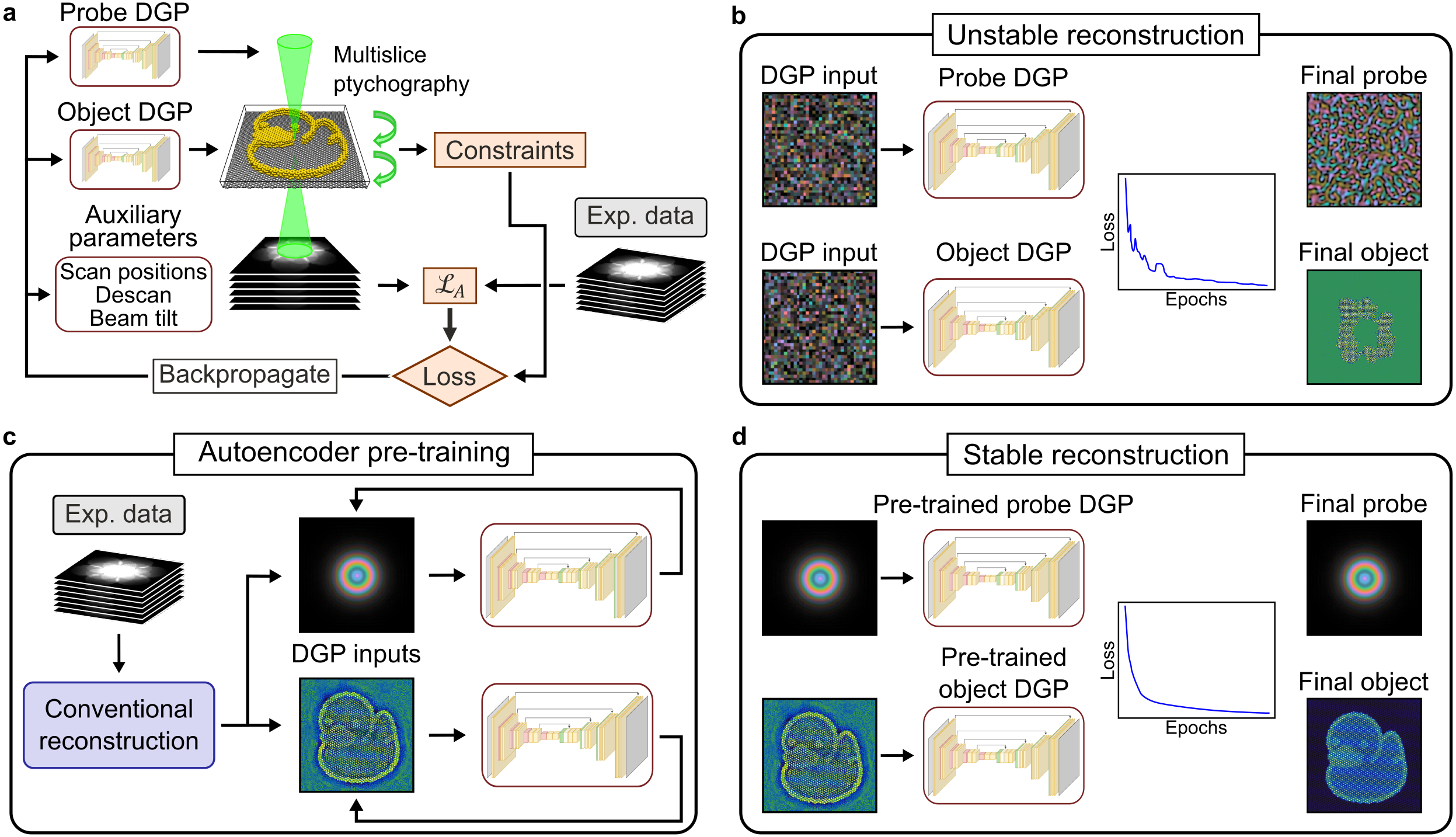}
\caption{
    \textbf{Overview of the DGP-enabled ptychographic reconstruction algorithm.} 
    \textbf{a} Two DGPs generate the sample object and probe used in a mixed-state multislice ptychography forward model. 
    Gradients are computed via automatic differentiation and backpropagated to simultaneously update both DGPs and any auxiliary parameters. 
    \textbf{b} Randomly instantiated DGPs with noise-only inputs often yield unstable reconstructions and rarely converge to a physical solution. 
    \textbf{c} The DGPs can be pre-trained as autoencoders using an approximate object and probe obtained from a conventional iterative or single-shot reconstruction. 
    \textbf{d} The pre-trained DGPs are then employed within the full ptychographic forward model to produce stable reconstructions. 
    }
\label{fig:outline}
\end{figure}

A flowchart of our ptychographic reconstruction algorithm is shown in Fig.~\ref{fig:outline}a. 
One DGP generates a complex-valued mixed-state electron probe, and a second DGP generates the 3D sample object. 
We use a mixed-state multislice electron ptychography forward model that includes learnable auxiliary parameters (e.g., scan positions). 
The forward model predicts diffraction intensities that are compared with experimental data to form a fidelity loss. 
Soft constraints, such as TV regularization, can be applied to the object or probe and contribute additional loss terms. 
Each component of the forward model is differentiable, enabling automatic differentiation and backpropagation to simultaneously optimize the probe DGP, object DGP, and auxiliary parameters.

The DGPs used in this work are 2D convolutional neural networks with a U-Net architecture \cite{ronneberger2015}. 
For 3D multislice reconstructions or mixed-state probes, the channel dimension indexes object slices along the beam direction or different probe modes. 
The shape of the DGP-generated object or probe is determined by the shape of the tensor input to the DGP. 
In typical DGP applications, such as image inpainting or super-resolution~\cite{ulyanov2018}, this input is random noise. 
However, we find that the noise-only input strategy can lead to unstable and nonphysical reconstructions, as illustrated in Fig.~\ref{fig:outline}b, due to simultaneously optimizing two DGPs through a complex forward model.
Figure \ref{fig:outline}c shows how this instability can be mitigated by pre-training the DGPs. 
First, we estimate the object and probe using a conventional direct or iterative pixelated reconstruction. 
We then train the object DGP as an autoencoder on the estimated object, by taking it as input and reproducing it as output, and we analogously train the probe DGP on the estimated probe. 
This pre-training is fast and stable because each model is trained independently without traversing the ptychographic forward model. 
The pre-trained models are subsequently used in the full ptychographic reconstruction (Fig.~\ref{fig:outline}d). 
We retain the estimated object and probe as inputs to each DGP so the reconstruction starts from a physically plausible state.

\subsection*{Noise robustness}
Using DGPs to generate the sample object and probe offers a number of advantages as opposed to learning a pixelated representation of either. 
The first benefit is an increased robustness to noisy data; a DGP leverages the regularizing nature of the CNN architecture to generate images and volumes with low patch-wise entropy~\cite{ulyanov2018}, producing less noisy reconstructions while reducing user-specified hyperparameters and regularizers. 

\begin{figure}[htb]
\centering
  \includegraphics[width=1.0 \linewidth]{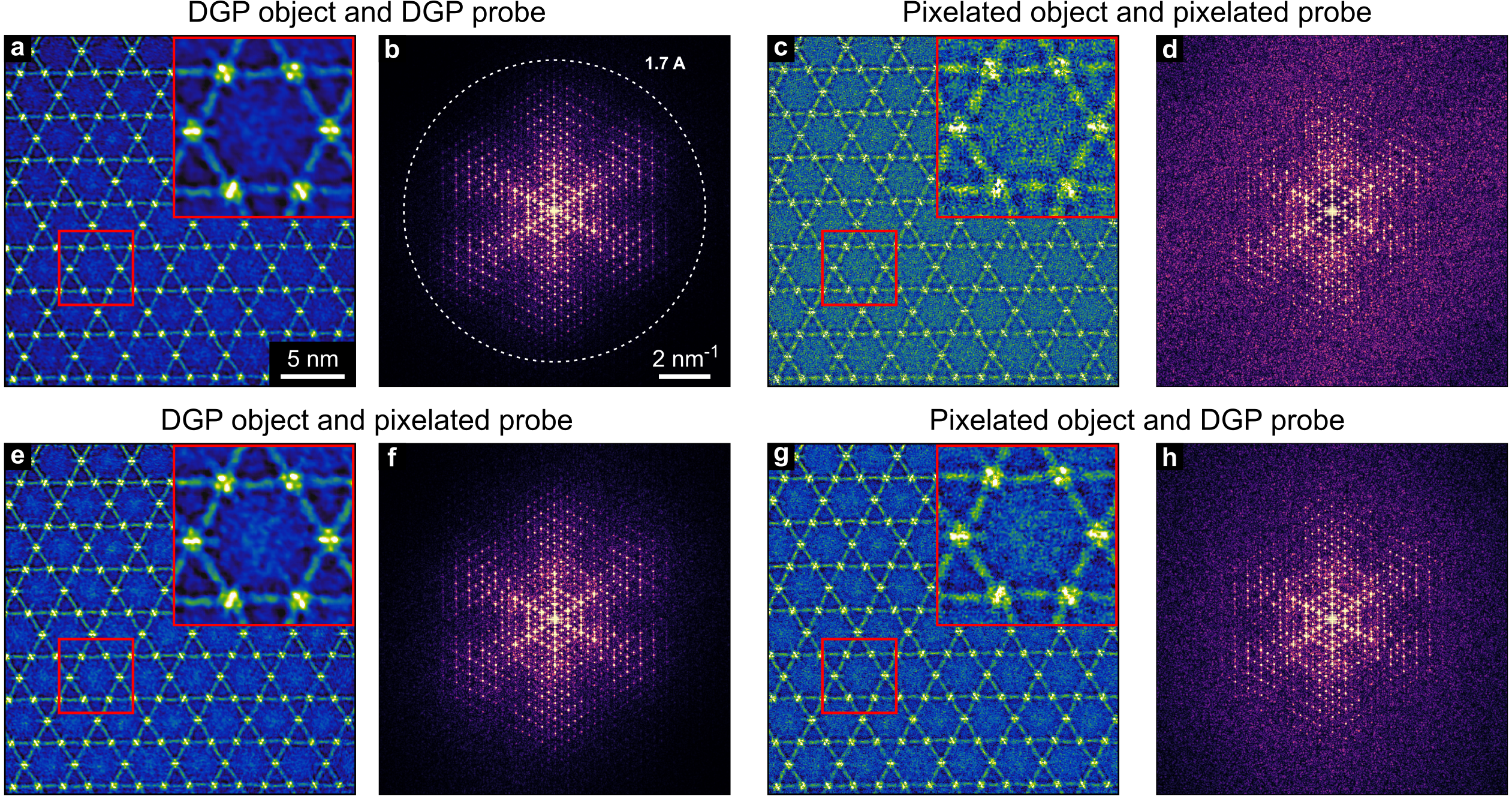}
  \caption{\textbf{Denoised reconstructions of the MOSS-6 MOF.} \textbf{a} Reconstruction using a DGP to generate both the sample object and probe with maximal resolution and minimal noise. \textbf{b} FFT of \textbf{a}. \textbf{c, d} Reconstruction and FFT of the same dataset using a pixelated object and probe. \textbf{e, f} Reconstruction and FFT using a DGP-generated object and a pixelated probe. \textbf{g, h} Reconstruction and FFT using a pixelated object and DGP-generated probe.}
  \label{fig:MOSS6}
\end{figure}

We demonstrate this by reconstructing a publicly available dataset from Li et al.~\cite{li2025} of the MOSS-6 metal-organic framework (MOF), imaged at \qty{100}{\electron \per \angstrom \squared}. 
Figure~\ref{fig:MOSS6}a, c, e, and g show the reconstructed object phase when using different combinations of pixelated and DGP-generated objects and probes. 
The best reconstruction uses DGPs for both the object and probe, as shown in Fig.~\ref{fig:MOSS6}a, which is much less noisy than the reconstruction shown in Fig.~\ref{fig:MOSS6}c that uses a pixelated object and probe. 
This is evident from the real-space images, where the \ce{Zr} atomic columns are more clearly defined in Fig.~\ref{fig:MOSS6}a, and it is especially visible in the fast Fourier transform (FFT) power spectra of the projected objects, which are shown in Fig.~\ref{fig:MOSS6}b and d. 
The DGP-based reconstruction reduces the noise profile across a wide range of frequencies without attenuating signal. 
Consequently, the information limit improves to \qty{1.57}{\angstrom}, versus \qty{1.98}{\angstrom} for the pixelated approach (Extended Data Fig.~\ref{figSup:moss6_info}). 

This noise reduction does not rely on additional regularization (beyond that provided by the DGPs) or filtering in the iterative reconstructions. 
All reconstructions use the same forward model and optimizers; the only difference is whether the object and/or probe are parameterized by DGPs or by pixelated grids. 
In all DGP-based reconstructions, the probe network was pretrained on a parameterized ideal probe model, while the object network required no pretraining.

To isolate each DGP’s effect, Fig.~\ref{fig:MOSS6}e and f show a reconstruction using a DGP-generated object and pixelated probe, and Fig.~\ref{fig:MOSS6}g and h show a reconstruction using a pixelated object and DGP-generated probe.
The probe DGP improves the reconstruction quality, and specifically leads to more isotropic information transfer, as seen by comparing Fig.~\ref{fig:MOSS6}d and h. 
The power spectrum in Fig.~\ref{fig:MOSS6}d is strongly asymmetric, and signal in the weaker directions is improved when using a probe DGP. 
It is clear from Fig.~\ref{fig:MOSS6}f that the object DGP is significantly more important for denoising the real-space reconstruction, which is expected because the object DGP can directly suppress noise. 
Nevertheless, in all cases the best reconstructions use DGPs for both object and probe.

\subsection*{Accelerated convergence}
Using DGPs in the ptychographic reconstruction algorithm can also substantially speed up convergence. 
Iterative ptychography algorithms have limited transfer of information for low spatial frequency signals, which can be problematic for samples containing nanoparticles, step edges, or holes~\cite{mccray2025, wittwer2022a}.
Such datasets require running the reconstruction algorithm for many iterations and therefore use hours of performant computational resources. 
This can significantly complicate the reconstruction process, as reconstructions are often run repeatedly to optimize hyperparameters and maximize image quality. 

\begin{figure}[htb]
\centering
  \includegraphics[width=1.0 \linewidth]{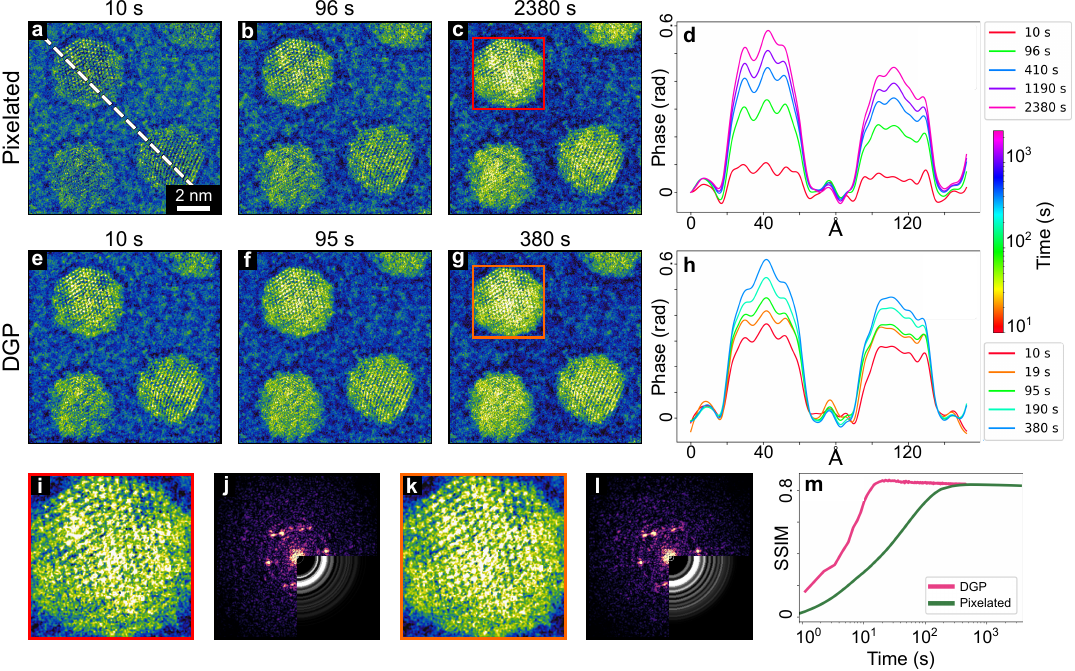}
  \caption{
  \textbf{Accelerating reconstructions of Au nanoparticles.} 
  \textbf{a}--\textbf{c} The reconstructed object after \qty{10}{\second}, \qty{96}{\second}, and \qty{2380}{\second} when using a pixelated object and probe. 
  \textbf{d} Line scan across the dashed line in \textbf{a} showing the slow reconstruction of the bulk nanoparticle phase shift. 
  \textbf{e}--\textbf{h} The same dataset reconstructed using a DGP for the object and probe. 
  \textbf{i} Magnified view of the red box in \textbf{c}. \textbf{j} FFT of \textbf{i}, inset shows the background subtracted radial profile of the FFT. 
  \textbf{k},\textbf{l} Magnified view and FFT of the orange box in \textbf{g}. 
  \textbf{m} Plotting the structural similarity vs time for pixelated and DGP reconstructions of a simulated Au nanoparticle dataset.
  } 
  \label{fig:AuNP}
\end{figure}

Figure \ref{fig:AuNP} demonstrates slow convergence by reconstructing a dataset of gold nanoparticles on an amorphous carbon substrate.
Snapshots from an AD-based reconstruction using a pixelated object and probe are shown in Fig.~\ref{fig:AuNP}a-c after \qty{10}{\second}, \qty{96}{\second}, and \qty{2380}{\second} respectively. 
The atomic columns of each nanoparticle are clearly defined after just \qty{10}{\second}, but the bulk shape function of each nanoparticle requires nearly 40 minutes to fully resolve. 
Line plots across the dashed line in Fig.~\ref{fig:AuNP}a are shown in Fig.~\ref{fig:AuNP}d and highlight the delayed recovery of low-frequency signal.
By contrast, using DGPs to generate the object and probe greatly accelerate the reconstruction, as shown in Fig.~\ref{fig:AuNP}e-h. 
The DGP-enabled reconstruction includes the bulk phase shifts from the first few iterations and fully converges in under 7 minutes.

The final reconstruction quality of the pixelated and DGP reconstructions are very similar. 
Figure \ref{fig:AuNP}i-l show magnified regions of a single nanoparticle and the associated power spectra for the two reconstructions. 
The real-space images look nearly identical, in contrast to what was presented with the MOSS-6 dataset in Fig.~\ref{fig:MOSS6}. 
This is expected, since the nanoparticle reconstruction is not dose-limited and the noise-reduction benefits of the DGPs are therefore less pronounced.
However, fitted radial profiles of the power spectra in Fig.~\ref{fig:AuNP}j and l show enhanced signal at high spatial frequencies in the DGP reconstruction, indicating a superior reconstruction.

When comparing the convergence rates of the DGP-enabled and pixelated reconstructions, we evaluate progress by wall-clock time rather than number of iterations.
One of the downsides of using DGPs is the increased computational cost per iteration, both in terms of GPU memory required and time. 
When using a pixelated object and probe, the gold nanoparticle data could be reconstructed at a rate of \qty{0.16}{\second \per iteration}. 
The DGP reconstruction, by contrast, ran at \qty{1.9}{\second \per iteration} using modestly sized CNNs (see Methods for details on model architecture). 
The slower reconstruction speed is offset by the efficiency of the DGPs, which require many fewer iterations to converge. 
The DGP reconstruction times reported in Fig.~\ref{fig:AuNP} exclude pre-training, a one-time cost not repeated during hyperparameter optimization.
For these reconstructions, the probe and object DGPs were each pre-trained for 50 iterations, with the estimated probe and object obtained from 50 iterations of a pixelated reconstruction. The total pre-training time was approximately \qty{10}{\second}, negligible relative to the total reconstruction time even when using DGPs. 

Simulations corroborate the speedup of low-frequency signals. We simulated a ptychography dataset of a single gold nanoparticle on amorphous carbon under similar imaging conditions, and Fig.~\ref{fig:AuNP}m plots the structural similarity (SSIM) score of reconstructed objects from both DGP and pixelated reconstructions as a function of time (reconstruction snapshots in Extended Data Fig.~\ref{figSup:simAuNP}). Consistent with experiment, the DGP-enabled reconstructions converge much faster than when using a pixelated object and probe. 

\subsection*{Depth regularization}
Ptychographic reconstructions are challenging due to the large number of degrees of freedom introduced by simultaneous optimization of the sample object and probe.
This is particularly true for multislice ptychography, where many distinct solutions can reduce the data-fidelity loss yet remain physically implausible. 
Multislice ptychography therefore highlights another benefit of using DGPs for ptychographic reconstruction, which is the inherent regularization provided by the CNN architecture. 
DGPs bias towards piecewise-smooth, structured reconstructions, which we find promotes sparsity and spatial coherence that aligns naturally with high-resolution imaging, including at atomic resolution. 
Using DGPs can thus yield more physical reconstructions while reducing the amount of explicit regularization a user must impose.

\begin{figure}[tbhp]
\centering
\includegraphics[width=0.95\linewidth]{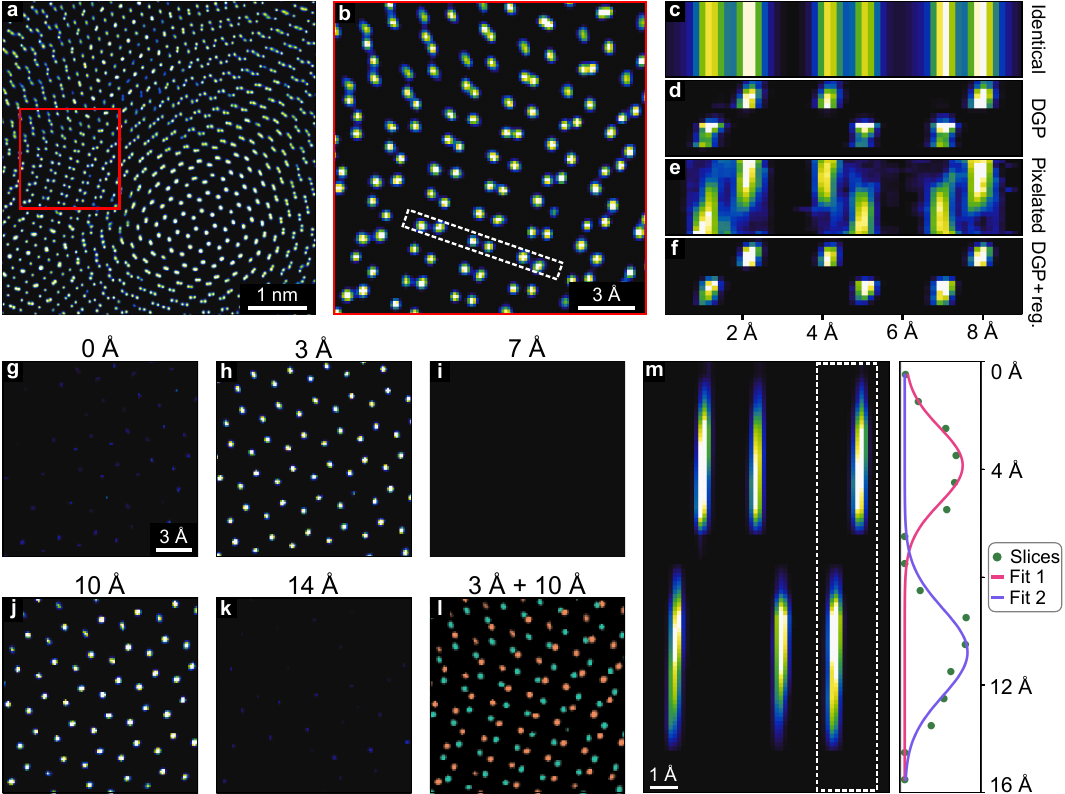}
\caption{
\textbf{Depth-regularization in a twisted bilayer of \ce{WSe2}.} 
\textbf{a} Full field-of-view reconstruction of two layers of \ce{WSe2} with a \qty{3}{\degree} twist angle. 
\textbf{b} Magnified view of the red box in \textbf{a}. 
\textbf{c}--\textbf{f} Cross-section view of the white boxed region in \textbf{b} for different reconstruction types and with different constraints applied: \textbf{c} DGP reconstruction with identical slices, \textbf{d} DGP reconstruction with free slices and no additional regularization, \textbf{e} pixelated reconstruction with free slices  and no additional regularization, \textbf{f} DGP reconstruction with free slices,  TV along the beam ($z$) direction, and a surface-zero loss. 
\textbf{g}--\textbf{k} Slices of the region shown in \textbf{b} for the DGP reconstruction with TV and surface-zero losses. 
\textbf{l} Overlay of \textbf{h} and \textbf{j} in green and orange, respectively. 
\textbf{m} Line profile shown in \textbf{f} rescaled so that the sampling is equal in $x$ and $y$. A double Gaussian is fit two the two atoms in the white box.
}
\label{fig:tBL}
\end{figure}

Fig.~\ref{fig:tBL} demonstrates DGP regularization for a twisted bilayer of \ce{WSe2} from Nguyen et al.~\cite{nguyen2024}. 
The phase reconstruction using two DGPs is shown for the full field-of-view and a magnified region in Fig.~\ref{fig:tBL}a and b.
The workflow had three stages: (i) a pixelated reconstruction for 50 iterations to produce object and probe estimates for DGP pretraining; (ii) a 50-iteration DGP run with an identical-slices constraint (uniform along the beam) to improve stability, allow higher learning rates, and accelerate convergence; and (iii) continued iterations after removing the identical-slices constraint, with or without additional regularization. 
We used a 16-slice multislice model with slice thickness \qty{1.0}{\angstrom}, and we enforced positivity of the reconstructed projected potential via a softplus final activation in the DGP networks.

Fig.~\ref{fig:tBL}c shows a cross-section of the boxed region in Fig.~\ref{fig:tBL}b after 50 iterations of DGP reconstruction with the identical slices constraint. 
After 20 unconstrained iterations (Fig.~\ref{fig:tBL}d), the DGP-based reconstruction separates the top and bottom \ce{WSe2} layers with minimal blending, unlike an identically run pixelated reconstruction (Fig.~\ref{fig:tBL}e).
Without any explicit regularization, the DGP can produce checkerboard artifacts along the beam direction, a known effect of convolutional architectures~\cite{odena2016}.  
We do not interpret the apparent splitting of atoms as the model correctly identifying stacked \ce{Se} atoms, because the splitting occurs nearly uniformly across all sites, including at the \ce{W} positions. 
This artifact is easily suppressed with a small amount of TV loss applied along the beam direction, yielding more physical atomic profiles.
We also include a ``surface-zero'' loss, which penalizes density on the top or bottom surfaces of the sample as described in the Methods. 
The result of using TV and surface-zero loss is shown in Fig.~\ref{fig:tBL}f.

The combination of DGP-enabled reconstruction with these two regularization terms yields two distinct layers of physically plausible atoms that are straightforward to interpret. 
Representative slices of the region highlighted in Fig.~\ref{fig:tBL}b are shown in Fig.~\ref{fig:tBL}g--k, emphasizing the clear separation between the \ce{WSe2} layers. 
Importantly, this method minimizes user bias by avoiding restrictive assumptions about atom shapes.
TV along the beam direction is used only to remove the minor checkerboard artifact; otherwise, the atom shapes and positions come entirely from the experimental data and the regularizing properties of the DGPs. 
Fig.~\ref{fig:tBL}m shows the cross-section from Fig.~\ref{fig:tBL}f rescaled to equal sampling in $x$ and $y$. 
Although atoms are less constrained along the beam, their profiles are physical. 
The interlayer spacing between the \ce{WSe2} sheets is \qty{7}{\angstrom}, consistent with the literature\cite{Xia2025}.

\subsection*{Effect of model depth}
When considering the benefits of using DGPs for ptychographic reconstruction, it is important to evaluate how different DGP architectures affect reconstruction quality. 
We have demonstrated that using DGPs relaxes the requirements for explicit regularizers and soft constraints, but this benefit would be limited if one must still extensively optimize hyperparameters in the form of the model architecture. 
Fortunately, we find that the DGP method is robust across different models, which we demonstrate by reconstructing the same dataset with object DGPs of varying depths.

\begin{figure}[htb]
\centering
  \includegraphics[width=1.0 \linewidth]{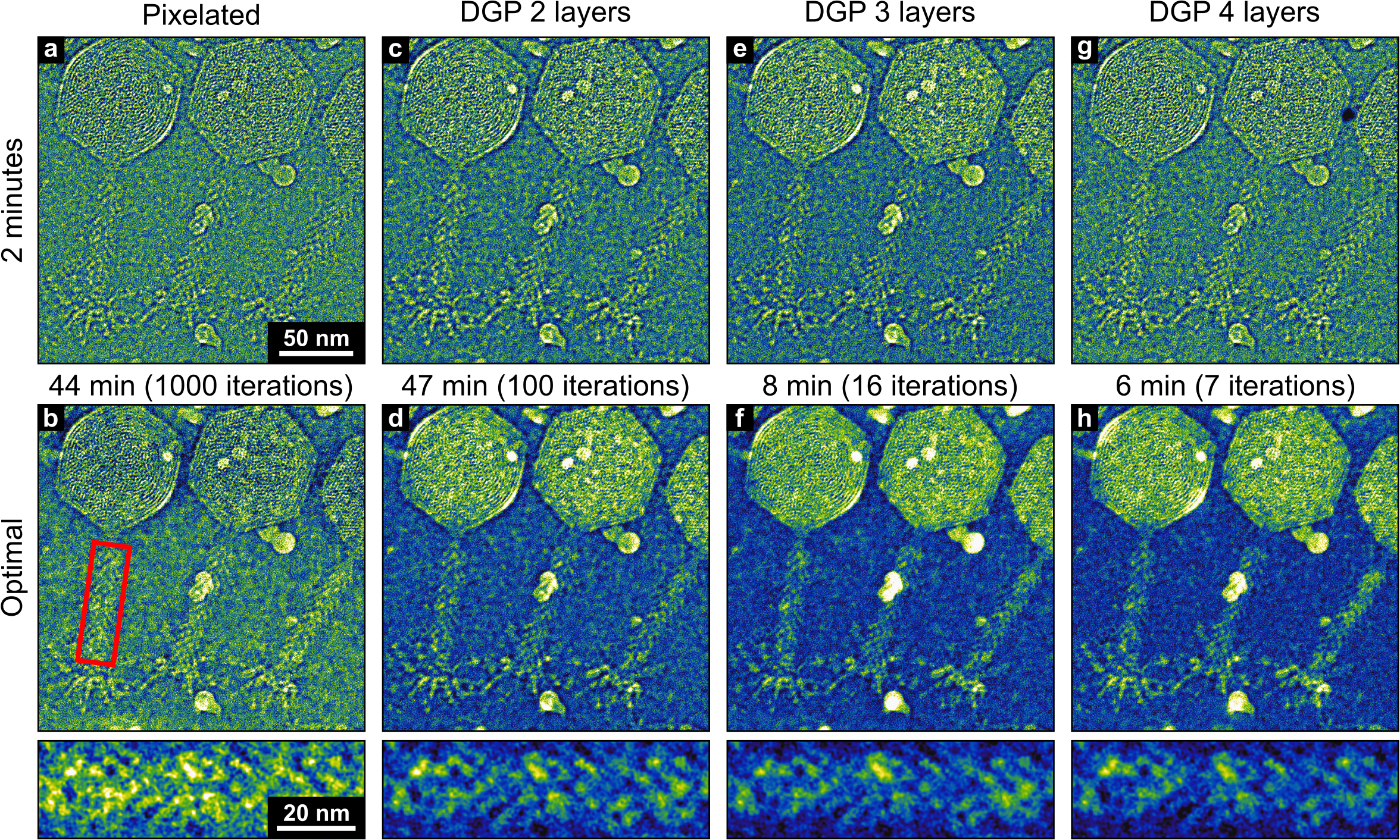}
  \caption{\textbf{Effects of DGP architecture when reconstructing biological data.} \textbf{a} Reconstructed object after \qty{2}{\minute} using a pixelated object and probe. \textbf{b} The optimal pixelated reconstruction was achieved after 1000 iterations and \qty{44}{\minute}. The lower image shows a magnified view of the red-boxed region highlighting the detail present in the bacteriophage sheath. \textbf{c} The same dataset after \qty{2}{\minute} reconstruction using a DGP object and probe. The object DGP is a 2-layer CNN. \textbf{d} The optimal reconstruction was achieved after 100 iterations and \qty{47}{\minute}. \textbf{e--h} The \qty{2}{\minute} and optimal reconstructions using a 3- and 4-layer CNN as the object DGP.} 
  \label{fig:virus}
\end{figure}

Fig.~\ref{fig:virus} shows multiple reconstructions of a Phi92 bacteriophage using a dataset published by K\"{u}\c{c}\"{u}ko\u{g}lu et al.~\cite{kucukoglu2024} that was acquired at \qty{49}{\electron \per \angstrom \squared}. 
A pixelated reconstruction after \qty{2}{\minute} is shown in Fig.~\ref{fig:virus}a, and the optimal reconstruction after 1000 iterations is shown in Fig.~\ref{fig:virus}e. 
We refer to this as the ``optimal'' reconstruction rather than the final or converged reconstruction because further iterations lead to overfitting to noise and artifacts in the dataset, which is apparent both from the appearance of artifacts in the reconstructed objects and supported by cross-validation (Extended Data Fig.~\ref{figSup:virus_pix}). 
A magnified view of the red-boxed region in Fig.~\ref{fig:virus}e is shown below each panel, highlighting structural details of the bacteriophage sheath.

Fig.~\ref{fig:virus}b--d show DGP reconstructions of the same dataset after \qty{2}{\minute} using object DGPs with different depths.
The 2-layer DGP produces a significantly improved final reconstruction compared to the pixelated method, as shown in Fig.~\ref{fig:virus}f, faithfully reproducing low-frequency signals in the object. 
However, it requires comparable computation time, reaching convergence in \qty{47}{\minute} over 1000 iterations. 
By contrast, the 3-layer DGP converged low spatial frequencies in only \qty{8}{\minute} and 16 iterations, while the 4-layer DGP converged in \qty{6}{\minute} and 7 iterations. 
All three DGP reconstructions achieve similar quality, though the 2-layer DGP does not fully capture the low-frequency background, and the 4-layer DGP exhibits stronger artifacts, such as those visible at the capsid edges. 
Additionally, the 4-layer CNN required more filters per layer than the shallower models, leading to substantially slower per-iteration computation (see Table~\ref{tab:DGP_stats}).

\begin{table}[htb]
    \centering
    \begin{tabular}{ccc}
        \# CNN layers & \# trainable parameters & speed (sec / iteration)\\
        \hline
        2 & \num{3.9e4} & \num{27.97}\\
        3 & \num{1.6e5} & \num{29.56}\\
        4 & \num{2.6e6} & \num{52.07}\\
    \end{tabular}
    \caption{Details of CNNs used as object DGPs in Fig.~\ref{fig:virus}.}
    \label{tab:DGP_stats}
\end{table}

Overfitting and the resulting artifacts represent the primary weakness of using deeper DGP models. 
Similar to pixelated reconstructions, deeper models can fit to artifacts and noise in experimental data, leading to unphysical reconstructions. 
Overfitting occurs more rapidly for deeper models, generally producing worse final results as illustrated by comparing Fig.~\ref{fig:virus}g and h. 
For the 4-layer DGP, artifacts appear by the time the low-frequency background has converged. 
By contrast, the shallower 2-layer DGP, while unable to fully reconstruct the low-frequency background, proved more resistant to overfitting; the reconstruction had effectively converged by 100 iterations, and further iterations produced minimal changes (Extended Data Fig.~\ref{figSup:virus_dgp}). 
These results suggest that moderately deep architectures (3 layers in this case) provide an optimal balance between convergence speed, reconstruction quality, and resistance to overfitting.

\section*{Discussion}
We have demonstrated that deep generative priors (DGPs) address key limitations of conventional ptychographic reconstruction algorithms across diverse experimental datasets. 
The key improvements, enhanced noise robustness, accelerated convergence, and physically plausible multislice reconstructions, stem from the implicit regularization inherent to convolutional neural network architectures, which naturally favor outputs with spatial coherence and characteristic low-frequency structure.

The noise-reduction capabilities vary with sample type in instructive ways. 
For the MOSS-6 MOF, DGPs significantly reduced high-frequency noise and improved the information limit, while the Phi92 bacteriophage showed less dramatic denoising. 
This difference arises because the DGP can learn and exploit repetitive structural motifs present in crystalline materials; the consistent unit cells and atomic basis enable the network to distinguish signal from noise more effectively than for aperiodic biological samples. 
Nevertheless, even for aperiodic samples, DGPs significantly accelerate convergence and reduce the need for careful tuning of regularization hyperparameters.
These results also suggest that reconstructing many biological datasets simultaneously with shared network weights could achieve better denoising.

A key practical consideration is the robustness of DGP-enabled reconstruction to model architecture choices. 
We find that sensible architectures, such as the 3-layer U-Net employed for most datasets, perform well across diverse sample types without extensive optimization. 
This robustness reduces the expertise barrier compared to the sensitivity of conventional methods to regularization weights and constraint parameters.

The pre-training strategy is essential for stable optimization when using DGPs for ptychography. 
Unlike typical applications of DGPs to single images, ptychographic reconstruction optimizes two networks simultaneously through a computationally expensive forward model. 
Pre-training each DGP as an autoencoder on approximate values from a short conventional reconstruction is a fast, one-time step that provides physically plausible starting points and dramatically improves convergence.

Comparing DGP-enabled ptychography to other machine learning approaches highlights distinct advantages. 
Supervised learning methods require large training datasets with known ground truth, which can be difficult to acquire and inherently limit application scope. 
By contrast, DGPs require no training data with known ground truths; regularization arises entirely from the network architecture. 
This makes DGP-enabled ptychography immediately applicable to new sample types and experimental conditions. 
The DGP framework is also fundamentally AD-based, retaining the flexibility to implement additional constraints and jointly optimize auxiliary parameters.

To facilitate community adoption, all methods developed in this work are available in the open-source Quantitative Electron Microscopy Python package (quantEM). 
QuantEM is designed to be accessible to non-computational experts while remaining highly modular, enabling application to diverse experimental conditions including segmented detectors and non-raster scans.
The reduced hyperparameter tuning required with this framework allows easier adoption, with less domain-specific knowledge to optimize reconstructions.
Future work will leverage implicit neural representations that make DGPs spatially-aware, enabling efficient distribution and reconstruction of large datasets across multiple GPUs. 
We anticipate that DGP-based electron ptychography will find broad application across the full range of materials and biological sciences.

\section*{Methods}
\subsection*{Computational methods}
The DGP-enabled ptychography code is implemented in the quantEM package and is built on PyTorch~\cite{ansel2024}. 
The DGPs are convolutional autoencoders, and the model inputs are the estimated probe and object that are obtained from either a direct reconstruction method or a few iterations of an AD-based pixelated reconstruction. 
A small amount of Gaussian noise ($\sigma=0.025$) is added to the DGP inputs each iteration in order to help the models avoid local minima. 
Noise is added to the original DGP input each time, i.e. the noise is not cumulative, and the noise is complex- or real-valued in accordance with the input datatype. 

For each DGP, hard constraints are applied directly to the model outputs, before the probe or object is used in the ptychography forward model. 
The only hard constraints used in this work were probe orthogonality and identical object slices, when applicable. 
The constrained object and probe are used in a mixed-state multislice ptychography forward model to obtain predicted diffraction patterns.
The forward model also includes auxiliary parameters so items like the precise scan positions can be optimized. 

The predicted diffraction patterns are compared with the experimental data to obtain a fidelity loss. 
Soft constraints, such as total variation (TV) and a surface-zero term, are calculated as additional loss terms that are added to the fidelity loss before backpropagation. 
The anisotropic TV loss takes the form of 
\begin{equation}
    \mathcal{L}_{TV}(V) = \frac{1}{XYZ}\sum_{i,j,k} \lambda_{xy} \left(|V_{i+1,j,k} - V_{i,j,k}| + |V_{i,j+1,k} - V_{i,j,k}|\right) + \lambda_z |V_{i,j,k+1} - V_{i,j,k}|,
\end{equation}
where $\lambda_{xy}$ and $\lambda_{z}$ are the TV weights for the in-plane and beam directions, and the object has shape $X \times Y \times Z$. 
For complex objects, the TV loss is calculated for the phase and amplitude independently and these are summed. 

The surface zero loss is computed as 
\begin{equation}
    \mathcal{L}_{surf}(V) = \frac{\lambda_{surf}}{2XY} \left( \sum_{i,j} |V_{i,j,0}| + \sum_{i,j} |V_{i,j,Z-1}| \right),
\end{equation}
where $\lambda_{surf}$ is the surface zero weight. 

The total loss is backpropagated and the gradients are calculated with reverse mode automatic differentiation. 
The DGPs each have their own optimizer and both DGPs are updated every reconstruction iteration. 
Precise implementation details, including example code and notebooks, are available in the \href{https://github.com/electronmicroscopy/quantem-tutorials}{quantEM tutorials} repository. 
All reconstructions presented in this work were run on an NVIDIA L40s GPU. 

\subsection*{DGP architecture}
The DGPs are fully-convolutional neural networks built on a U-Net model as depicted in Extended Data Fig.~\ref{figSup:UNET}~\cite{ronneberger2015}. 
The default architecture, for both object and probe DGPs, was a 3-layer CNN with 16 starting filters. 
The number of input and output channels corresponds to the number of probe modes for the probe DGP and the number of object slices for the object DGP. 
The DGPs used a rectified linear unit (ReLU) activation function, except for the final layer which used an identity function. 
When reconstructing a potential object for the \ce{WSe2} data in Fig.~\ref{fig:tBL}, positivity was enforced in the object with a softplus final activation instead. 

The probe DGP is always complex valued, but the object DGP can be complex- or real-valued, depending on the object type. 
Complex-valued objects are reconstructed using a complex-valued DGP, but if reconstructing a pure-phase or potential object, a real-valued DGP is used.

\section*{Code Availability}
They Python code developed in this study is available in the open-source \href{https://github.com/electronmicroscopy/quantem}{Quantitative Electron Microscopy (quantEM)} package. Tutorial notebooks and demonstrations reproducing these results are available in the accompanying \href{https://github.com/electronmicroscopy/quantem-tutorials}{quantEM tutorials} repository. 

\section*{Data Availability}
The experimental MOSS-6, \ce{WSe2}, and Phi92 data are available from the original publications \cite{li2025, nguyen2024, kucukoglu2024}. The gold nanoparticle data is available through the \href{https://github.com/electronmicroscopy/quantem-tutorials}{quantEM tutorials} repository.

\bibliography{references}

\section*{Acknowledgments}
Work at the Molecular Foundry was supported by the Office of Science, Office of Basic Energy Sciences, of the U.S. Department of Energy under Contract No. DE-AC02-05CH11231.

\section*{Competing Interests}
The authors declare no competing interests.

\section*{Additional information}
Extended data for this paper is available at <link>.

\clearpage

\renewcommand{\figurename}{Extended Data Fig.}
\setcounter{figure}{0}

\begin{figure}[htb]
\centering
  \includegraphics[width=0.95\linewidth]{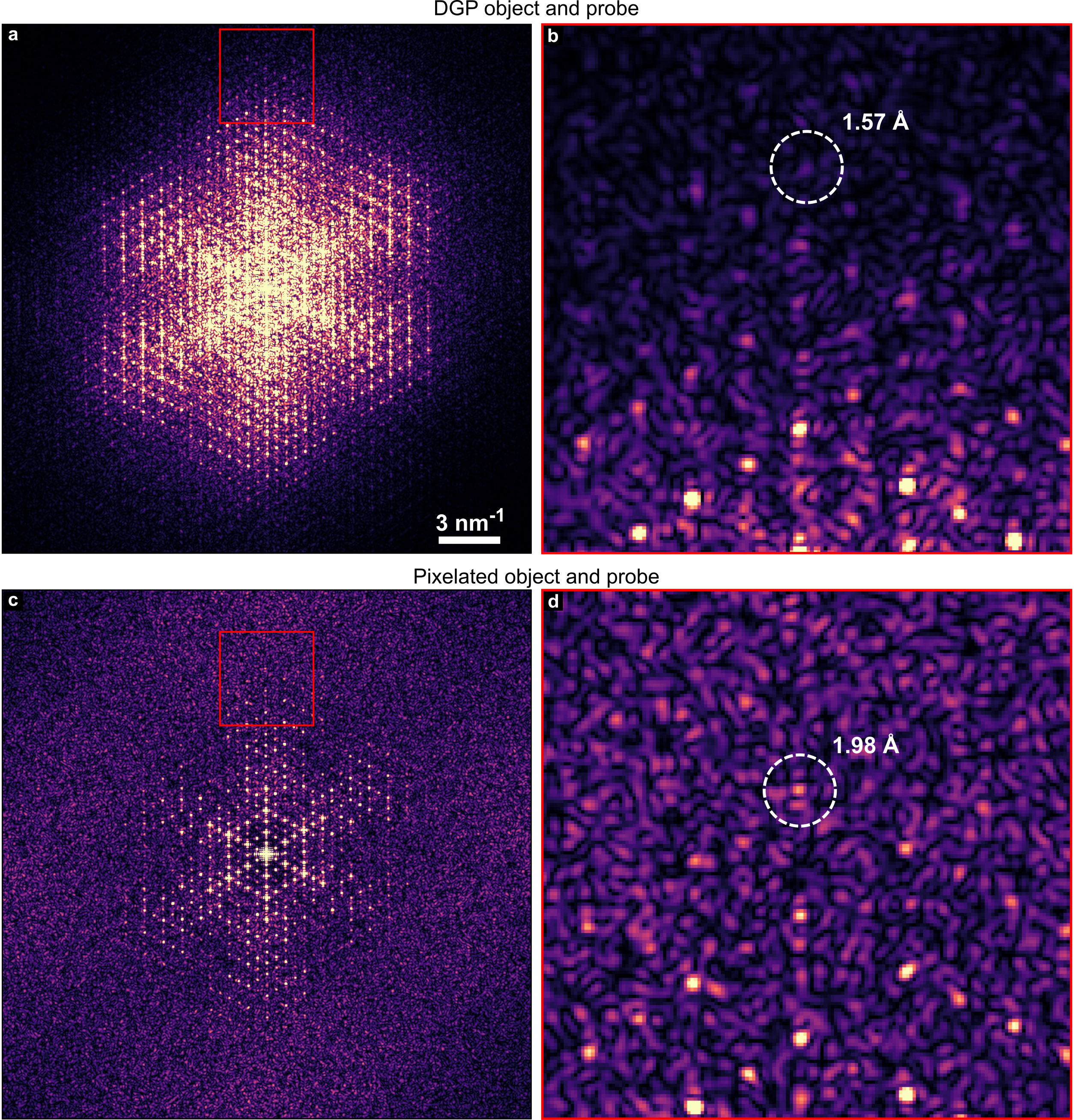}
  \caption{\textbf{Information limit for the MOSS-6 reconstructions.} \textbf{a} Power spectrum of the reconstructed object using a DGP to generate the object and probe. This is the same data as in main text Fig.~\ref{fig:MOSS6}b, but shown here with a lower contrast range to emphasize the weaker, high-frequency signal. \textbf{b} Magnified view of the red boxed region in \textbf{a}, with a circle identifying a peak at \qty{1.57}{\angstrom}. \textbf{c} Same as \textbf{a} but for the pixelated reconstruction as shown in Fig.~\ref{fig:MOSS6}d. \textbf{d} The identified peak from \textbf{c} showing an information limit of \qty{1.98}{\angstrom}. 
  }
  \label{figSup:moss6_info}
\end{figure}

\begin{figure}[htb]
\centering
  \includegraphics[width=0.9\linewidth]{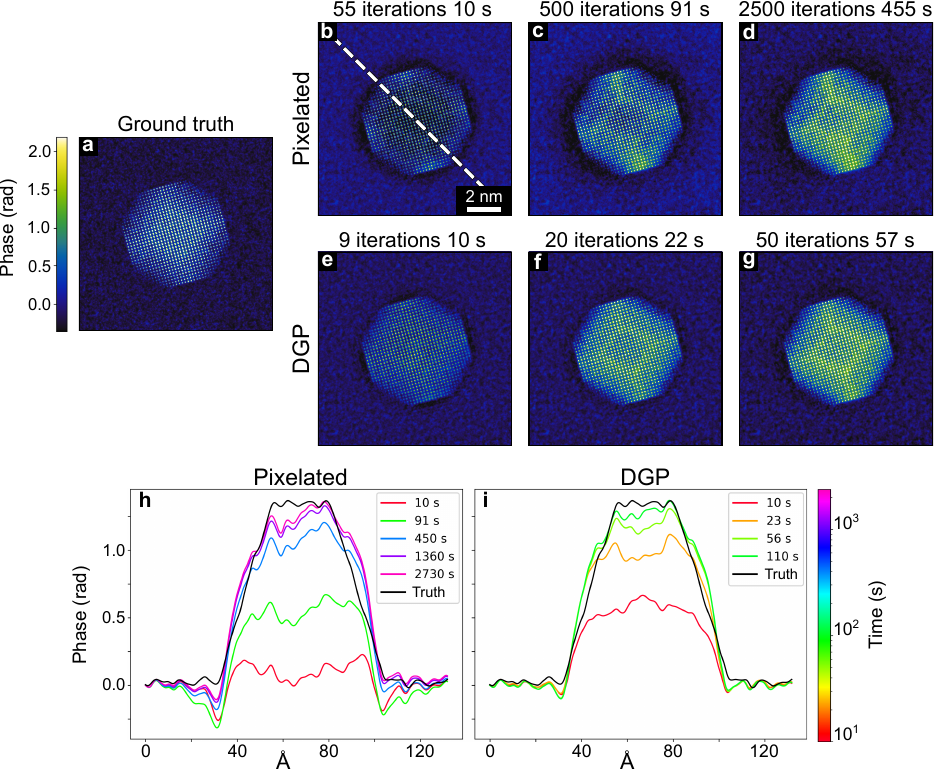}
  \caption{\textbf{Reconstructing a simulated gold nanoparticle.} \textbf{a} The ground truth phase shift from a simulated gold nanoparticle on amorphous carbon. \textbf{b--d} Snapshots of the reconstruction using a pixelated object and probe, after \qty{10}{\second}, \qty{91}{\second}, and \qty{455}{\second} respectively. \textbf{e--g} Snapshots of the reconstructed object using a DGP-generated object and probe, after \qty{10}{\second}, \qty{22}{\second}, and \qty{57}{\second} respectively. \textbf{h, i} Line plots across the dashed line in \textbf{b} showing the convergence of the bulk nanoparticle phase shift for the pixelated and DGP reconstructions. 
  }
  \label{figSup:simAuNP}
\end{figure}

\begin{figure}[htb]
\centering
  \includegraphics[width=1.0\linewidth]{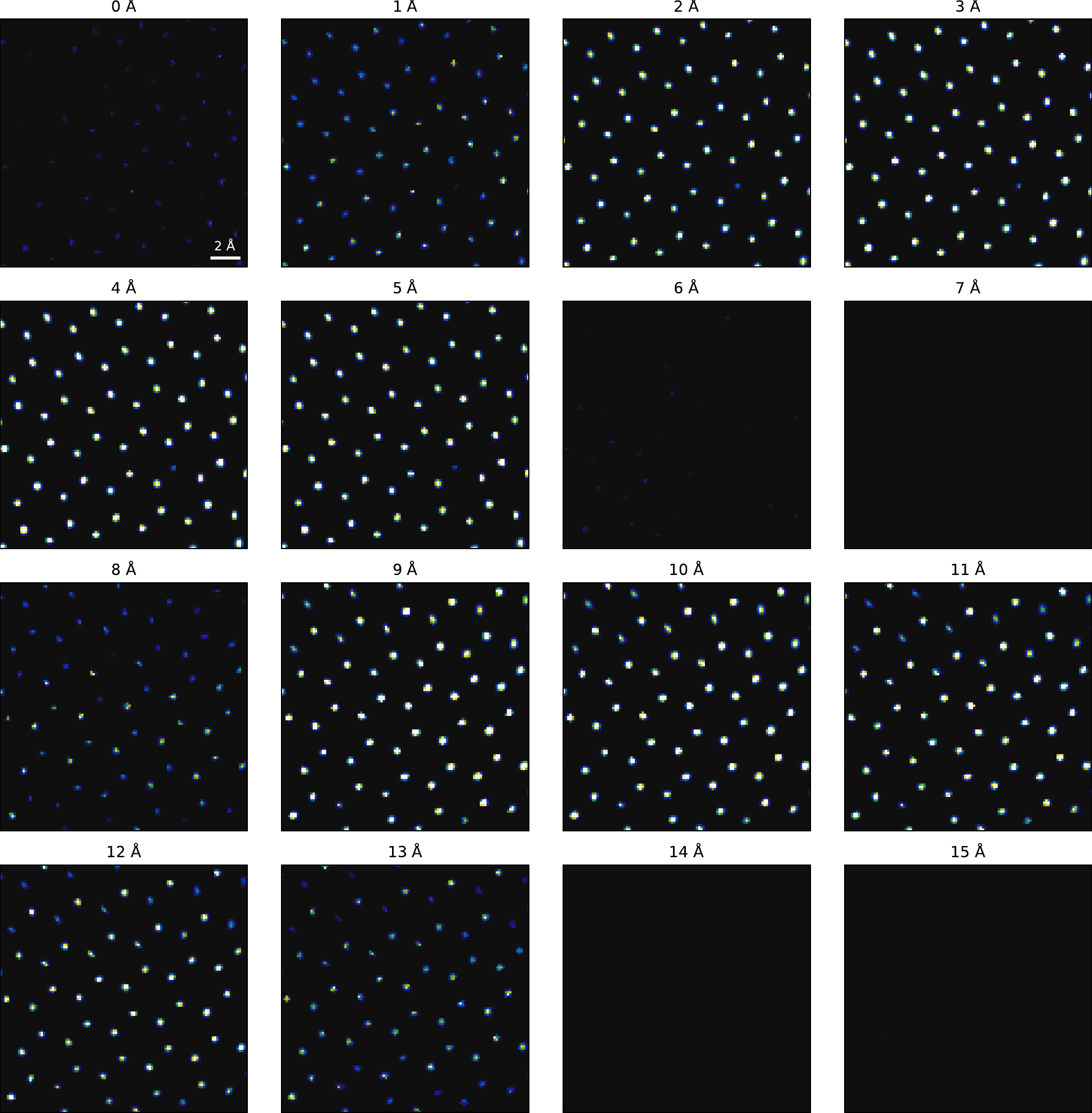}
  \caption{\textbf{DGP multislice reconstruction of twisted bilayer \ce{WSe2}.} All slices are shown with the same contrast range. Selected slices are also shown in main text Fig.~\ref{fig:tBL}g--k. 
  }
  \label{figSup:DGP_tBL_slices}
\end{figure}

\begin{figure}[htb]
\centering
  \includegraphics[width=0.9\linewidth]{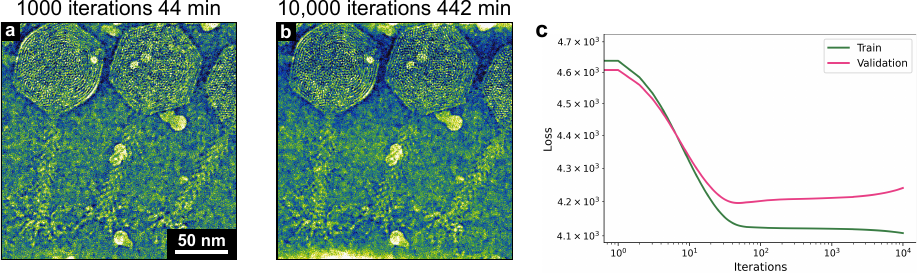}
  \caption{\textbf{Effect of additional reconstruction iterations.} \textbf{a} Pixelated reconstruction of the Phi92 bacteriophage after 1000 iterations, which was deemed the ``optimal'' reconstruction and was presented in main text Fig.~\ref{fig:virus}e. \textbf{b} Additional iterations do not improve the reconstruction, and low-frequency artifacts are more apparent. \textbf{c} Training and cross-validation loss using \qty{10}{\percent} of the probe positions as validation data. The increasing validation loss implies that the model reconstruction is overfitting. 
  }
  \label{figSup:virus_pix}
\end{figure}

\begin{figure}[htb]
\centering
  \includegraphics[width=0.95\linewidth]{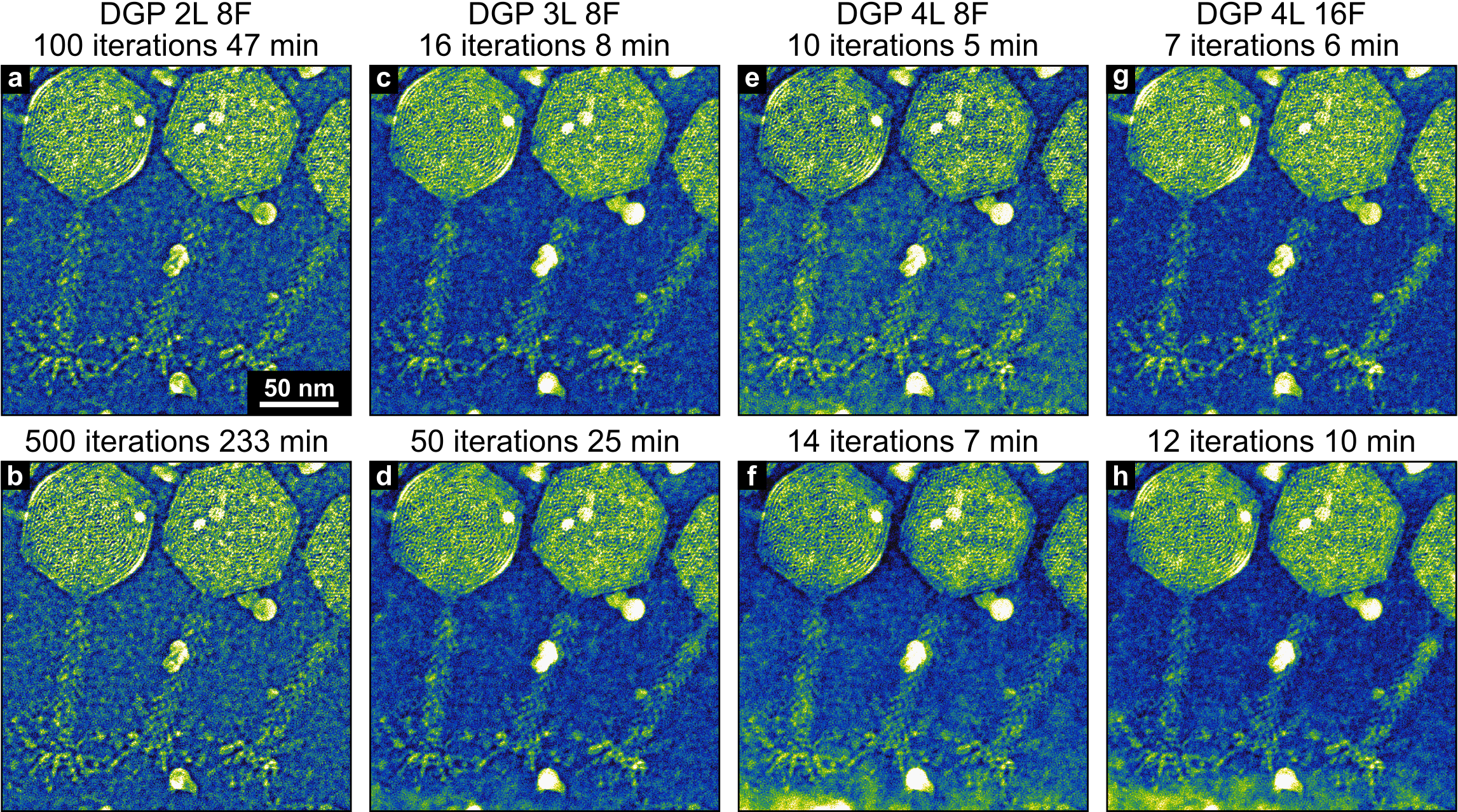}
  \caption{\textbf{Effects of DGP architecture and additional reconstruction iterations.} \textbf{a} Optimal DGP reconstruction using a 2-layer CNN with a starting filter size of $\mathrm{F}=8$ as the object DGP, reproduced from main text Fig.~\ref{fig:virus}f. \textbf{b} Additional iterations does not improve or significantly change the reconstruction when using a shallow DGP. \textbf{c} Using a 3-layer CNN as the object DGP leads to a better reconstruction and faster convergence. \textbf{d} Additional iterations with the 3-layer DGP lead to overfitting, as seen at the bottom edge of the image. \textbf{e} Using a 4-layer CNN as the DGP with $\mathrm{F}=8$, as was used for the 2- and 3-layer DGPs, does not lead to a good reconstruction. The model overfits before the low-frequency background is learned. \textbf{f} Reconstruction after 14 iterations showing how the 4-layer DGP quickly overfits the data. \textbf{g} Increasing the number of starting filters for the 4-layer DGP to $\mathrm{F}=16$ significantly improves the reconstruction quality, at the cost of significantly increasing the computation time per iteration. \textbf{h} The larger DGP is quicker to overfit the data than when using shallower models. 
  }
  \label{figSup:virus_dgp}
\end{figure}

\begin{figure}[htb]
\centering
  \includegraphics[width=0.8\linewidth]{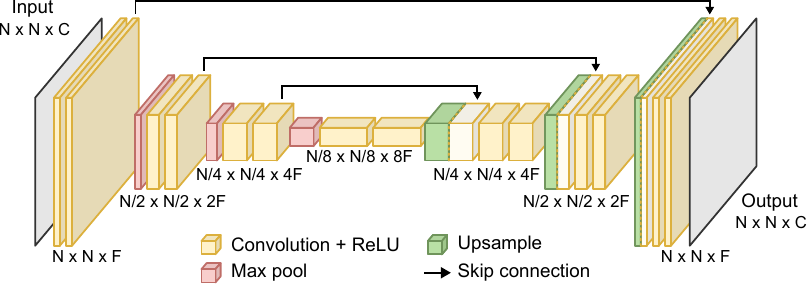}
  \caption{\textbf{Block diagram of a 3-layer CNNs used as a DGP.} The DGPs used for this work are fully convolutional neural networks based on a U-Net architecture. The number of input/output channels $\mathrm{C}$ corresponds to the number of probe modes for the probe DGP and the number of object slices for the object DGP. With the exception of main text Fig.~\ref{fig:virus}, all DGPs had 3 layers and the starting number of filters, $\mathrm{F}$, was 16. 
  }
  \label{figSup:UNET}
\end{figure}

\end{document}